\documentclass[pra,longbibliography,twocolumn,showpacs,nofootinbib,superscriptaddress,notitlepage,accepted=2017-12-23]{quantumarticle}
\usepackage{graphicx}
\usepackage{amsmath}
\usepackage{amssymb,bm}
\usepackage{color,dsfont}
\usepackage[colorlinks=true, hyperindex, breaklinks, linkcolor=blue, urlcolor=blue, citecolor=blue]{hyperref} 
\usepackage[numbers,sort&compress]{natbib}
\usepackage[normalem]{ulem}
\usepackage[capitalise]{cleveref}
\usepackage{mathrsfs}
\usepackage{subcaption}
\usepackage{ragged2e}
\DeclareCaptionJustification{justified}{\justifying}
\usepackage{mathtools}
\usepackage{lipsum}
\usepackage{float}

\newcommand{\cE}{\mathcal{E}}
\newcommand{\cS}{\mathcal{S}}

\newcommand{\cJ}{\mathcal{J}}
\newcommand{\cP}{\mathcal{P}}
\newcommand{\cT}{\mathcal{T}}

\newcommand{\ket}[1]{|#1\rangle}  
 
\newcommand{\trace}[1]{\mathrm{Tr}\left(#1\right)}

\begin{document}

\title{Fault-tolerant quantum computing in the Pauli or Clifford frame with slow error diagnostics}

\author{Christopher Chamberland}
\email{c6chambe@uwaterloo.ca}
\affiliation{
    Institute for Quantum Computing and Department of Physics and Astronomy,
    University of Waterloo,
    Waterloo, Ontario, N2L 3G1, Canada
    }
\author{Pavithran Iyer}
\email{pavithran.iyer.sridharan@usherbrooke.ca}
\affiliation{
   D\'epartement de Physique and Institut Quantique, 
   Universit\'e de Sherbrooke,
   Sherbrooke, Qu\'ebec, J1K 2R1 Canada
    }
\author{David Poulin}
\email{david.poulin@usherbrooke.ca}
\affiliation{
    D\'epartement de Physique and Institut Quantique, 
    Universit\'e de Sherbrooke,
    Sherbrooke, Qu\'ebec, J1K 2R1 Canada
    }

\begin{abstract}
We consider the problem of fault-tolerant quantum computation in the presence of slow error diagnostics, either caused by  measurement latencies or slow decoding algorithms. Our scheme offers a few improvements over previously existing solutions, for instance it does not require active error correction and results in a reduced error-correction overhead when error diagnostics is much slower than the gate time. In addition, we adapt our protocol to cases where the underlying error correction strategy chooses the optimal correction amongst all Clifford gates instead of the usual Pauli gates. The resulting Clifford frame protocol is of independent interest as it can increase error thresholds and could find applications in other areas of quantum computation.
\end{abstract}


\maketitle

\section{Introduction and Motivation}
\label{sec:Intro}

In fault-tolerant quantum computation, syndrome measurements are used to detect and diagnose errors. Once diagnosed, an error can be corrected before it propagates through the rest of the computation. In this article, we are concerned with the impact of slow error diagnostics on fault-tolerance schemes. There are two origins for this concern. First, in certain solid-state  and ion-trap qubit architectures, measurement times can be between 10 to 1000 times slower than gates times \cite{BKM+14,JSM1500,PJT+05,VHY+14,SLQEHP15,OHMMMYM09,HRB08}. Thus, on the natural operating time-scale of the quantum computer, there is a long delay between an error event and its detection. Second, processing the measurement data to diagnose an error---i.e., decoding---can be computationally demanding. Thus, there can be an additional delay between the data acquisition and the error identification. At first glance, these delays might cause the error probability to build up between logical gates, thus effectively decreasing the fault-tolerance threshold. But as we will see, this is not necessarily the case.

One of the key tricks to cope with slow error diagnostics is the use of error frames \cite{Knill05}. While the basic idea of error correction is to diagnose and correct errors, it can often be more efficient to keep track of the correction in classical software instead of performing active error correction. In particular, this saves us from performing additional gates on the system, thus removing potential sources of errors. In most quantum error-correction schemes, the correction consists of a Pauli operator, i.e., tensor products of the identity $I$ and the three Pauli matrices $X,\ Y,$ and $Z$. Thus, at any given time, the computation is operated in a random but known {\em Pauli frame}: its state at time $t$ is $P\ket{\psi(t)}$  for some Pauli operator $P$ and where $\ket{\psi(t)}$ denotes the ideal state at time $t$. When error-diagnostics is slow, the system will unavoidably evolve in an unknown error frame for some time.

The problem of slow measurements in solid-state systems was addressed by DiVincenzo and Aliferis \cite{DA07} in the context of concatenated codes. In addition to error diagnostics, concatenated schemes require measurements to prepare certain ancilla states used to fault-tolerantly extract the error syndrome and to inject magic states  to complete a universal gate set. DiVincenzo and Aliferis' scheme hinges on the fact that the logical gate rate decreases exponentially with the number of concatenation levels; thus, at a sufficiently high level, measurement and gate times become commensurate. To concretely realize this simple observation, they combine a number of known and new techniques including ancilla correction, high-level state injection, and Pauli frames. 

One limitation of this solution is that it only applies to concatenated codes realizing a universal gate set through magic state injection. This  leaves out for instance surface codes \cite{FMMC12} or concatenated codes with alternate universal gate constructions \cite{KLZ96,JL14,PR13,ADP14,Bombin14,YTC16}. In particular, they inject noisy magic-states directly at high concatenation levels, thus losing the benefit of low-level correction. In addition, when decoding times are very slow, this solution could wastefully use additional layers of concatenation with the sole purpose of slowing down the logical gates. Another drawback of their scheme is that it requires active error correction to ensure that high-level corrections are always trivial. 

In this article, we introduce an alternative scheme to cope with slow error diagnostics which applies broadly to all codes and universal gate constructions.  Like the DiVincenzo and Aliferis scheme, the key idea will be to slow-down the logical gate rate to learn the error frame before it propagates to the rest of the computation. This is simply achieved by spacing out gates in the logical circuit and thus circumvents the unnecessary additional qubit overhead associated to extra concatenation layers. Our scheme does not require active error correction, it is entirely realized in an error frame. In addition, our scheme is compatible with a more general form of error correction which uses {\em Clifford frames}, where at any given time $t$, the state of the computer is $C\ket{\psi(t)}$ where $C$ is a tensor-product of single-qubit Clifford group elements (defined below). Note that details of the Pauli frame scheme are given in \cref{sec:PauliFrame} whereas details of the Clifford frame scheme are given in \cref{sec:CliffordFrame}.

Regarding slow measurements, it is important to distinguish two time scales. First, the time $t_l$ it takes for the outcome of a measurement to become accessible to the  outside world.  For instance, $t_l$ could be caused by various amplification stages of the measurement, and we refer to it as a measurement {\em latency}. Second, the measurement {\em repetition} time $t_r$ is the minimum time between consecutive measurements of a given qubit. In principle, $t_r$ can be made as small as the two-qubit gate time, provided that a large pool of fresh qubits are accessible. Indeed, it is possible to measure a qubit by performing a CNOT to an ancillary qubit initially prepared in the state $\ket 0$ and then measuring this ancillary qubit. While the ancillary qubit  may be held back by measurement  latencies for a time $t_l$ before it can be reset and used again, other fresh qubits can be brought in to perform more measurements in the meantime. The scheme we present here and the one presented in \cite{DA07} are designed to cope with measurement latencies $t_l$, but both require small $t_r$. Indeed, the accuracy threshold is a function of the error rate per gate time (including measurement gates). Thus, increasing the measurement repetition time $t_r$ relative to the decoherence rate will unavoidably yield a lower threshold.

While the DiVincenzo and Aliferis scheme was motivated by slow measurements, most of it applies directly to the case of slow decoding. Our scheme too is oblivious to the origin of slow error diagnostics. We emphasize that slow decoding is a very serious concern. For instance, numerical simulations used to estimate the threshold or overhead of the surface code are computationally dominated by the decoding algorithm and require intense computational resources. Depending on the code distance and error rate, a single decoding cycle can take well above 1$\mu s$ \cite{DFTMN10} on a standard processor, considerably slower than the natural GHz gate rate in the solid state. 

As explained in \cite{Barbara15}, if the rate of the classical syndrome processing (decoding) is smaller than the rate at which the syndrome is generated, an exponential slow down would occur during the computation. However, just as measurements with long latencies can be handled with a supply of additional fresh measurement qubits, slow decoding can be handled with a supply of parallel classical computers so that the overall decoding rate matches the syndrome creation rate. Thus, in this article, slow error diagnostics is used to designate \textit{latencies} in measurements and/or decoding. 

Finally, to our knowledge, a theory of Clifford frames for fault-tolerant quantum computation has not been worked out before. By enabling them, our scheme offers a greater flexibility for error correction, thus potentially correcting previously non-correctable error models, or increasing the threshold of other noise models \cite{CWBL16}. More generally, the possibility of fault-tolerantly computing in a Clifford frame opens up the door to new fault-tolerant protocols, e.g.,  using new codes or new hardware that have a different set of native fault-tolerant gates. For instance,  Clifford frames were used as an accessory in measurement-based quantum computation with Majorana fermions \cite{KKL16}. Lastly, the possibility of computing in a Clifford frame could have applications to randomized compiling \cite{WE16} which introduces random frames to de-correlate physical errors.

\section{Pauli frame}
\label{sec:PauliFrame}

For the remainder of the paper, we will often make use of the following definitions. First, we define $\cP_n^{(1)}$ to be the $n$-qubit Pauli group (i.e. $n$-fold tensor products of the identity $I$ and the three Pauli matrices $X,\ Y,$ and $Z$). Then the {\em Clifford hierarchy} is defined by $\cP_n^{(k)} = \{U : UPU^\dagger \in \cP_n^{(k-1)}\quad \forall P \in \cP_n^{(1)}\}$. 

In physical implementations where measurement times are much longer than gate times or when error decoding is slow, if one were to perform active error correction, a large number of errors could potentially build up in memory during the readout times of the measurement. However, for circuits containing only Clifford gates, all Pauli recovery operators can be tracked in classical software without ever exiting the Pauli group. Indeed, suppose that at some given time during the computation, the state of the computer is in some Pauli frame defined by $P$---i.e., the state of the computer is $P\ket\psi$ where $\ket\psi$ is the ideal state (here $P$ can be any element of $\cP_n^{(1)}$, not necessarily a logical Pauli operator). If we then apply a {\em Clifford gate} $U\in \cP_n^{(2)}$ (again, not necessarily a logical gate), then the state will be $UP\ket\psi = P'U\ket\psi$ where $P'$ is another Pauli operator. We thus see that the effect of $U$ is to correctly transform the perfect state $\ket\psi$ and to change the Pauli frame $P$ in some known way. Moreover, updating the Pauli frame $P' = UPU^\dagger$ can be done efficiently, with complexity $\mathcal{O}(n^2)$ \cite{Gottesman98}. This shows that as long as we only apply Clifford gates, there is no need to actively error-correct, we can instead efficiently keep track of the Pauli frame in classical software \cite{Knill05}. 

\begin{figure}
\centering
\includegraphics[width=0.35\textwidth]{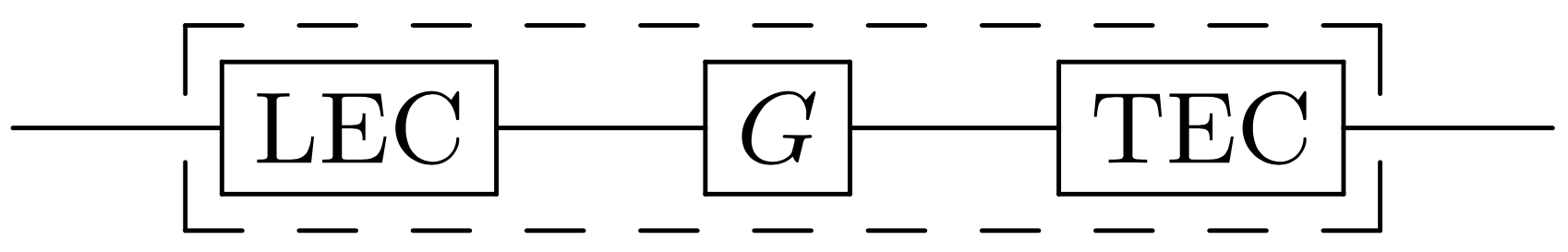}
\caption{Example of extended rectangle (exRec) for implementing the logical gate $G$. The leading and trailing EC circuits (LEC and TEC) perform fault-tolerant error correction for input errors and errors occurring during the implementation of $G$.}
\label{fig:exRec}
\end{figure}

In concatenated codes for instance, error correction is performed in between the application of gates to ensure that errors don't accumulate in an uncontrolled fashion. A gate location in a quantum algorithm is thus replaced by an extended rectangle, where error correction is performed both before and after the application of the gate \cite{AGP06}. The leading error correction circuit in an extended rectangle is used to correct input errors that could have occurred prior to the application of the gate. The trailing error correction circuits correct errors that can occur during the application of the gate (see \cref{fig:exRec}). Each error correction circuit will multiply the current Pauli frame by a Pauli operator (the correction).

Since Clifford gates can be efficiently simulated on a classical computer, non-Clifford gates are needed for universal quantum computation. Universal quantum computation can be achieved for instance using gates from the Clifford group combined with the $T =  \mathrm{diag}(1,e^{i\pi/4}) \in \cP_1^{(3)}$ gate \cite{BMPRV99}. In general, Pauli operators will not remain in the Pauli-group when conjugated by non-Clifford gates and so the Pauli frame cannot simply propagate through them. 

Consider the application of a logical $T$ gate in a quantum algorithm. Since measurement times are much longer than gate times, the Pauli frame right before the application of a $T$ gate can only be known at a later time. Furthermore, by definition of the Clifford hierarchy, Pauli operators are mapped to Clifford operators under conjugation by $T\in \cP_1^{(3)}$ gates. Therefore, the output correction after applying a logical $T$ gate can be outside the Pauli frame.

In order to overcome these obstacles, we note that if error correction is performed immediately after the application of the $T$ gate, the output correction can be written as a {\em logical} Clifford gate $C$ times a Pauli matrix (a proof is presented in \cref{app:CliffordAndPureError}).

\begin{figure}[h]
\includegraphics[scale=0.22]{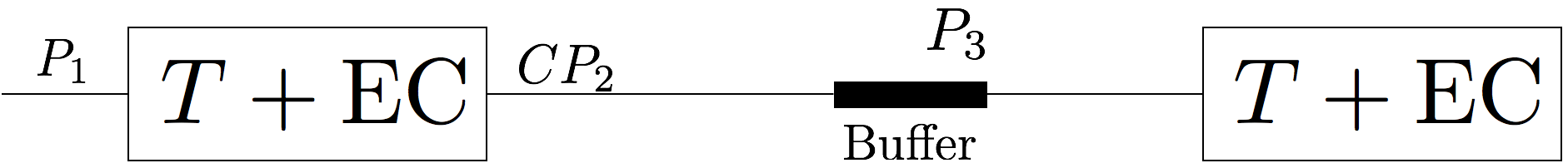} 
\vspace{0.4cm}

\includegraphics[scale=0.205]{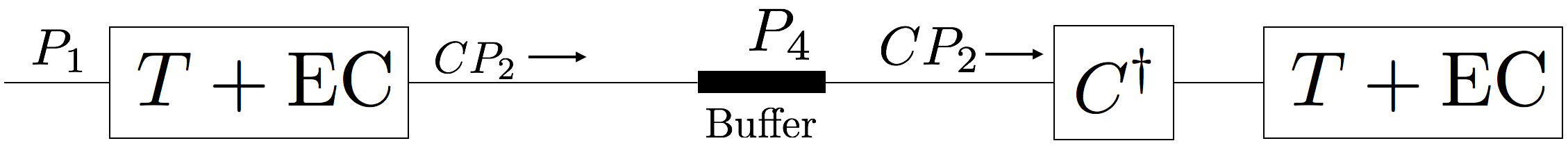}
\caption{Illustration of the scheme for propagating Pauli corrections through a $T$ gate when error diagnostics are much longer than gate times. (TOP)  When propagating the input Pauli $P_{1}$ through a $T$ gate and performing error correction, the output can be written as $CP_{2}$ where $C$ is a logical Clifford gate and $P_{2}$ is a Pauli matrix. A buffer is introduced to learn the Pauli frame immediately before applying the $T$ gate which enables the logical Clifford correction $C$ to be known. During the buffer, repeated rounds of error correction are performed to prevent the accumulation of errors for qubits waiting in memory. We denote the final Pauli correction arising from the EC rounds as $P_{3}$. (BOTTOM) We propagate the correction $CP_{2}$ through the buffer and apply a logical Clifford gate $C^{\dagger}$ in order to remove $C$ thus restoring the Pauli frame. Although the propagation can map the buffer correction $P_{3} \to P_{4}$, $P_{4}$ remains Pauli and can be known at a later time.}
\label{fig:PauliFrameScheme}
\end{figure}

If we were to keep track of the logical Clifford correction $C$ in classical software, it could propagate through the next $T$ gate, resulting in a correction involving even more $T$ gates. It could also propagate through a logical two-qubit gate (such as a CNOT) resulting in a two-qubit correction (the exact transformation rules are derived in \cref{sec:CliffordFrame}). The two-qubit corrections could then propagate through other gates in the quantum algorithm leading to a generic Clifford correction. To prevent these scenarios from occurring, a buffer can be inserted right before the next logical two-qubit gate or $T$ gate part of the quantum algorithm. The role of the buffer is to learn what the Pauli frame was right before the application of the previous logical $T$ gate. During the buffer, repeated rounds of error correction are performed to prevent the accumulation of a large number of errors. There could be leftover Pauli corrections arising from error correction rounds which would only be known at a later time. However, by propagating the logical Clifford correction through the buffer, the Pauli corrections would remain Pauli.

\begin{figure}
\centering
\includegraphics[width=0.4\textwidth]{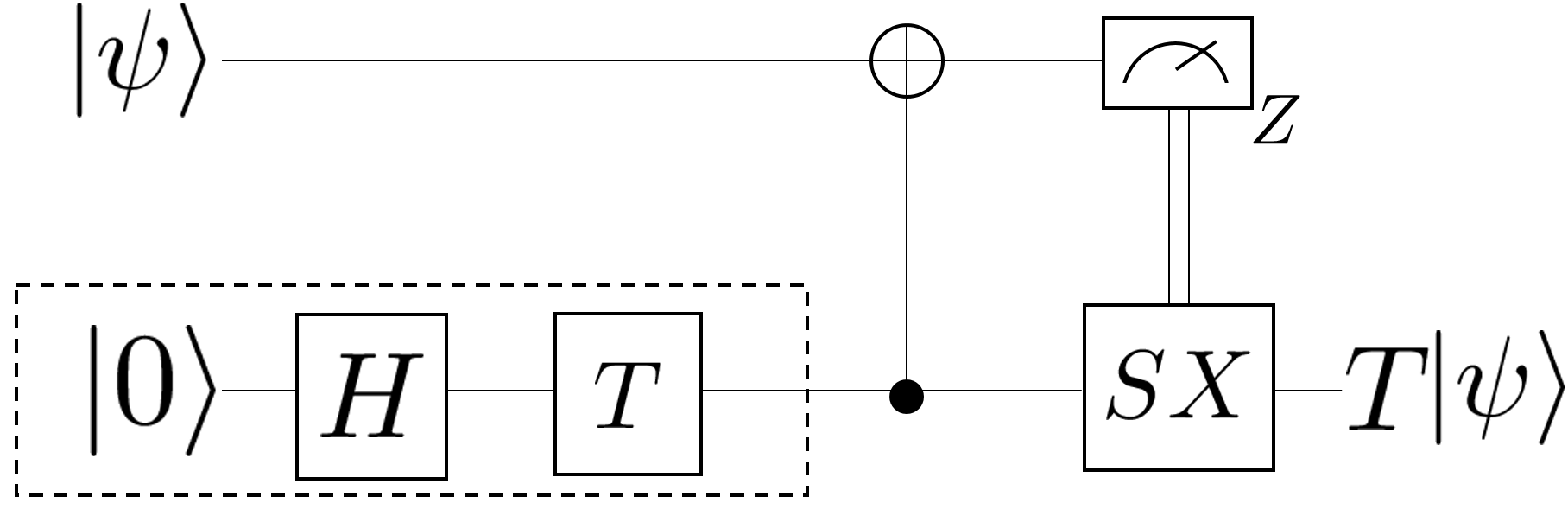}
\caption{Circuit for implementing a $T$ gate. The circuit uses the state $T\ket{+}$ which is prepared offline and applies a sequence of Clifford gates. The correction $SX$ is only applied if the $Z$-basis measurement outcome is -1.}
\label{fig:TgateStateInjection}
\end{figure}

Once the logical Clifford correction is propagated through the buffer, we apply a logical $C^{\dagger}$ thus restoring the Pauli frame. The protocol guarantees that only Pauli corrections would be propagated through the next logical $T$ or two-qubit gates. An illustration of the scheme is outlined in \cref{fig:PauliFrameScheme}. As in \cite{DA07}, the proposed approach also effectively slows down gate times making them comparable to measurement times\footnote{Note that the buffer increases the overall time for implementing a non-Clifford gate. However, this impacts the overall running time of the quantum algorithm only by a constant factor.}. However, buffers are only introduced when necessary and without having to increase the size of the code.   
                              
We conclude this section by noting that in \cite{DA07,Barbara15}, the Pauli frame scheme was described in the context of concatenated codes where $T$ gates are implemented using state injection as shown in \cref{fig:TgateStateInjection}. In these schemes, a buffer is included to learn what the logical Pauli frame was before applying the $SX$ correction in order to correctly interpret the outcome of the $Z$-basis measurement. For instance, if a logical Pauli $X$ occurred on the data prior to implementing the $T$ gate, the Clifford correction $SX$ would be applied if the measurement outcome was $+1$ instead of $-1$ (see the caption of \cref{fig:TgateStateInjection}). Further, as was described in \cite{DA07},  additional layers of concatenation are required to prevent the build up of errors in the presence of quantum measurements with a long latency.

While it builds on the same general ideas, the Pauli frame scheme presented in this section is not restricted to a particular implementation of the $T$ gate (for instance, it also directly applies to codes which can implement a logical $T$ gate transversely) nor to concatenated schemes. In fact, our scheme slightly differs from \cite{DA07,Barbara15} even in the case where the $T$ gate is applied using state injection. Indeed, one does not wait to learn what the frame was before determining whether to apply the logical $SX$ correction (the $Z$-basis measurement outcome is always interpreted in the same way). The entire state injection circuit is completed before the Pauli frame (prior to the application of the state injection circuit) is known. Once it becomes known, one would know if the wrong logical $SX$ correction was applied and any remaining logical Clifford errors would be removed. So in particular, there is no need to introduce additional coding layers to slow-down the logical gate rate during this waiting time. While this is a fairly minor distinction, it does enable us to generalize to any other coding schemes and implementations of the $T$ gate, a feature which has not been addressed prior to our work.

\section{Clifford frame}
\label{sec:CliffordFrame}

As was shown in \cite{CWBL16}, including Clifford corrections as part of the recovery protocol for error correcting codes that can implement logical Clifford gates transversely could significantly improve the code's threshold for certain noise models. In fact, for coherent noise channels ($\mathcal{N}(\rho)=e^{i\boldsymbol{n}\cdot \boldsymbol{\sigma}}\rho e^{-i\boldsymbol{n}\cdot \boldsymbol{\sigma}}$ where $\boldsymbol{n}=(n_{x},n_{y},n_{z})$ with $||\boldsymbol{n}||=1$), it was shown that in some regimes the 5-qubit code (see \cite{LMPZ96}) can tolerate an arbitrary amount of coherent noise when Clifford corrections are used, while Pauli corrections have a finite threshold.

In this section, we extend the protocol used to cope with slow error-diagnostics to the case where error correction uses Clifford gates. When performing error correction on a set of encoded qubits with a stabilizer code, if one measures a non-trivial syndrome value $l$, a recovery map $R_{l}$ is applied to the data block. The recovery map can always be written in the form $R_{l} = \mathcal{L}({l})\mathcal{T}({l})\mathcal{G}({l})$ where $\mathcal{G}({l}) \in \cP_n^{(1)}$ is a product of stabilizer generators, $\mathcal{L}({l})$ is a product of logical operators and $\mathcal{T}({l})\in \cP_n^{(1)}$ is a product of pure errors \cite{Poulin06}, see also \cref{app:CliffordAndPureError}. Pure errors form an abelian group of operators that commute with all of the code's logical operators and all but one of the codes stabilizer generators. The logical operators $\mathcal{L}({l})$ are chosen to maximize the probability of recovering the correct codeword -- this is the decoding problem. The operators in the set $\mathcal{L}({l})$ are not necessarily restricted to logical Pauli operators as they can be extended to include all logical operators that can be applied fault-tolerantly. 

It is thus natural to restrict $\mathcal{L}({l})$ to gates that can be applied transversally on the code. In particular, we will consider logical corrections in $(\cP_1^{(2)})^{\otimes n}$, the group generated by $n$-fold tensor products of single-qubit Clifford operators. This latter group is generated by
\begin{align}
H = \frac{1}{\sqrt{2}} \left( \begin{array}{cc}
                                          1 & 1 \\
                                          1 &-1\\                                                                     
                                          \end{array} \right),
\ \ {\rm and}\ \  
S = \frac{1}{\sqrt{2}} \left( \begin{array}{cc}
                                          1 & 0 \\
                                          0 & i\\                                                                     
                                          \end{array} \right),
\label{eq:HOp}
\end{align}
i.e., $\mathcal{P}^{(2)}_{1} = \langle H,S \rangle$. The order of the group is $|\mathcal{P}^{(2)}_{1}| = 24$ (ignoring global phases). For the $n$-qubit case, $\mathcal{P}^{(2)}_{n} = \langle H_{i},S_{i},\mathrm{CNOT}_{ij} \rangle$ where the indices $i,j$ indicate which qubits to apply the Clifford gates. We point out that 2-D color codes \cite{Bombin14} admit transversal realizations of all gates in $\cP_1^{(2)}$. Furthermore, the 5-qubit code \cite{BDSW96, LMPZ96} admits transversal realizations of logical gates in the set generated by $\langle SH,X,Z \rangle$.

For concatenated codes, we restrict the discussion to the case where logical Clifford corrections are performed at the last concatenation level only. If Clifford corrections were performed at every level, one would need to wait for the measurement outcomes of every level and actively perform Clifford corrections. This is because a level-$k$ logical Clifford correction does not generally commute with the level-($k+1$) syndrome measurements. The goal of defining a Clifford frame is to avoid actively correcting since the corrections themselves can introduce more errors into the computation. 

We now derive the transformation rules for Clifford operators propagating through CNOT gates. Two-qubit controlled unitary gates $C\mbox{-}U_{12}|a\rangle |b \rangle = |a\rangle U^{a} |b \rangle$, where the first qubit is the control and the second qubit is the target, can be written as
\begin{align}
C\mbox{-}U_{12} = \frac{1}{2}(I+Z)\otimes I + \frac{1}{2}(I-Z)\otimes U.
\label{eq:CNOT12}
\end{align}
Note that from this definition $\mathrm{CNOT}_{12} = C\mbox{-}X_{12}$. Using \cref{eq:CNOT12}, it is straightforward to show the following relations
\begin{align}
(S\otimes I)C\mbox{-}X_{12} &= C\mbox{-}X_{12}(S\otimes I),
\label{eq:PhaseControl}\\
(I\otimes S)C\mbox{-}X_{12} &= C\mbox{-}Y_{12}(I\otimes S),
\label{eq:PhaseTarget}\\
(I\otimes H)C\mbox{-}X_{12} &= C\mbox{-}Z_{12}(I\otimes H),
\label{eq:HadamardTarget}\\
(H\otimes I)C\mbox{-}X_{12} &= U_{X}(H\otimes I),
\label{eq:HadamardControl}
\end{align}
where we defined $U_{X} \equiv  \frac{1}{2}(I+X)\otimes I + \frac{1}{2}(I-X)\otimes X$.

\begin{figure}[h]
\centering
\begin{subfigure}{0.5\textwidth}
\includegraphics[width=\textwidth]{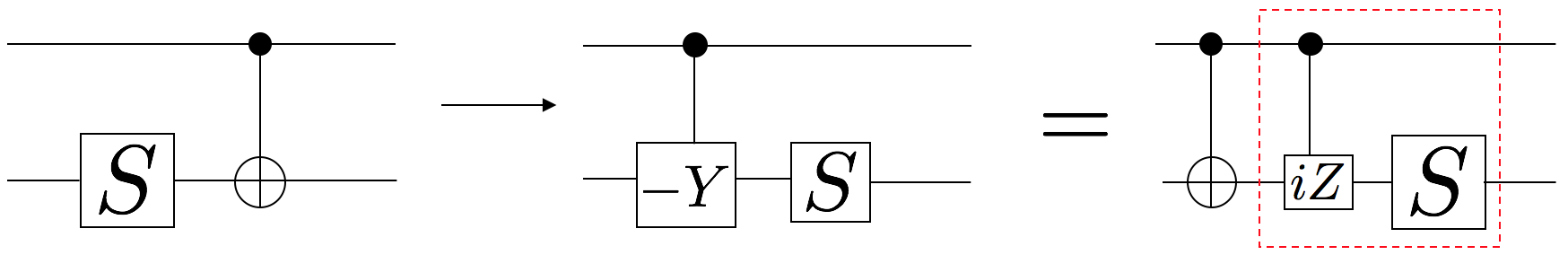}
\caption{}
\label{fig:PhaseTransformDiagram}
\end{subfigure}
\begin{subfigure}{0.5\textwidth}
\includegraphics[width=\textwidth]{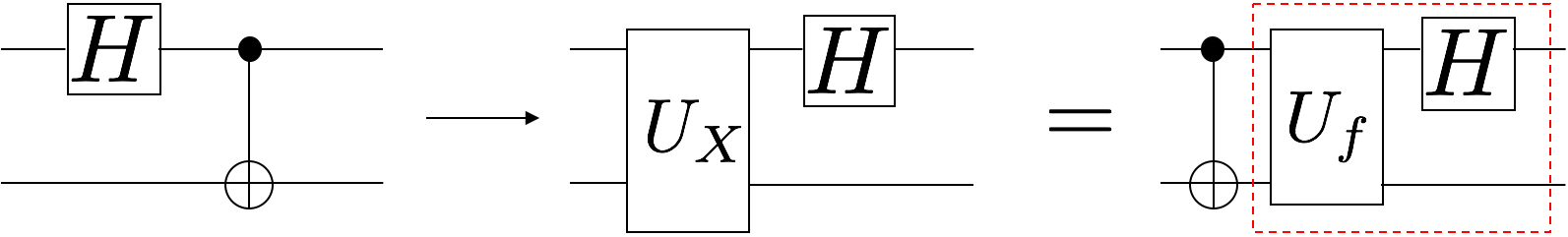}
\caption{}
\label{fig:HadTransformControlDiagram}
\end{subfigure}
\caption{(a) Propagation of $I\otimes S$ through a $\mathrm{CNOT}$ gate. (b) Propagation of $H \otimes I$ through a $\mathrm{CNOT}$ gate where $U_{f} = \frac{1}{2}(I+iY)\otimes I + \frac{1}{2}(I-iY)\otimes X$. In both cases, if instead of correcting the Clifford errors prior to the $\mathrm{CNOT}$ gate we were to keep track of them using a Clifford frame, the corrections would involve two-qubit gates in addition to the original Clifford corrections.}
\label{fig:PhaseTransDiag}
\end{figure}

From \cref{fig:PhaseTransDiag} and \cref{eq:PhaseControl,eq:PhaseTarget,eq:HadamardTarget,eq:HadamardControl}, it can be seen that propagating Clifford corrections through CNOT gates can result in both single and two-qubit Clifford corrections.  By keeping track of logical Clifford corrections in classical software, these could grow due to other CNOT gates resulting in a generic Clifford correction. Furthermore, when propagating Clifford corrections through non-Clifford gates (such as the $T$ gate), the output can potentially be outside of the Clifford hierarchy. These could then propagate through the remainder of the logical circuit resulting in a unitary gate correction expressed as a product of several $T$ gates. 

\subsection{Clifford propagation through CNOT gates}
\label{subsubsec:CliffPropCNOT}

\begin{figure}[h]
\centering
\begin{subfigure}{0.4\textwidth}
\includegraphics[width=\textwidth]{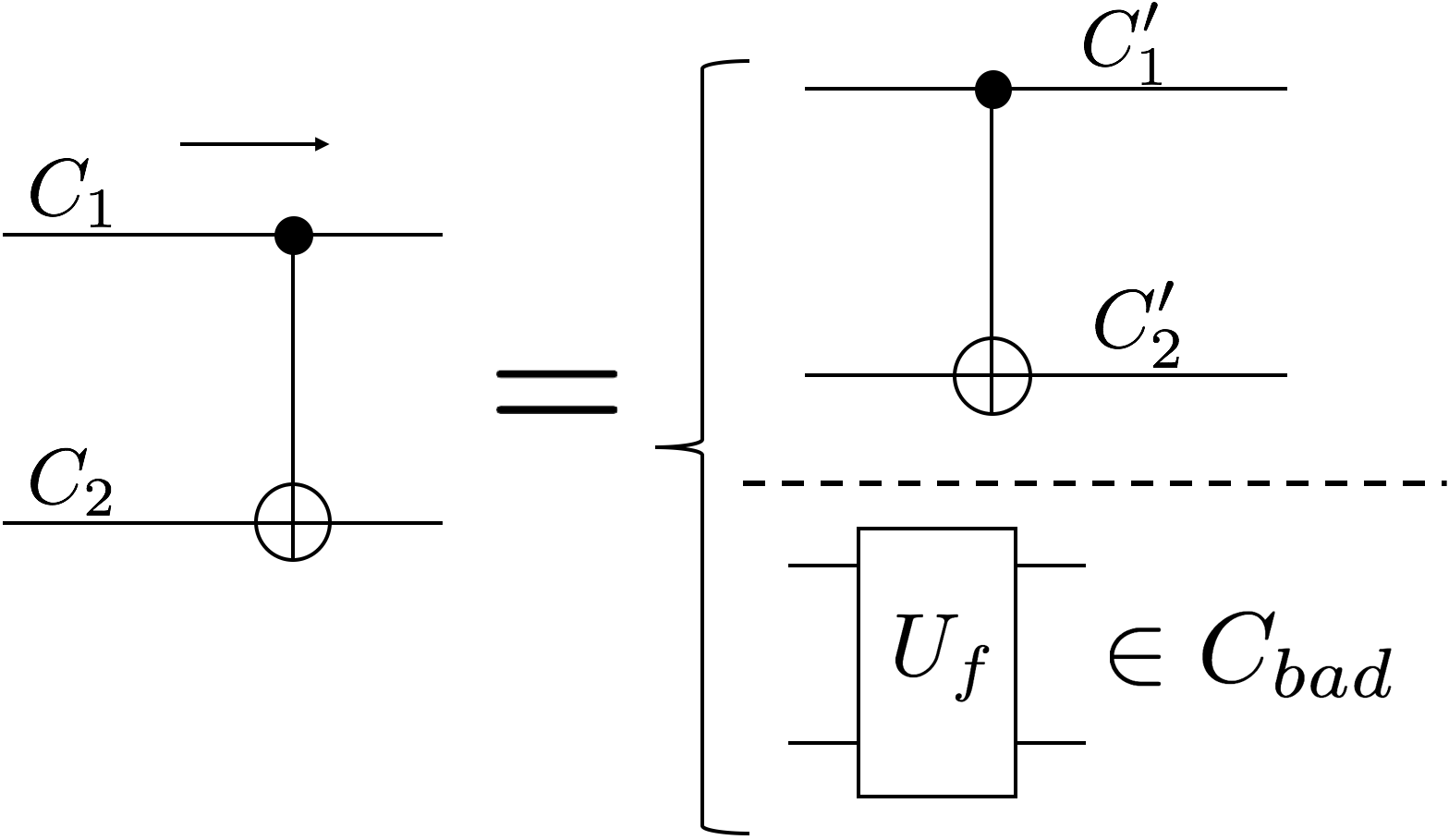}
\caption{}
\label{fig:CNOTtensorProp1}
\end{subfigure}
\begin{subfigure}{0.30\textwidth}
\includegraphics[width=\textwidth]{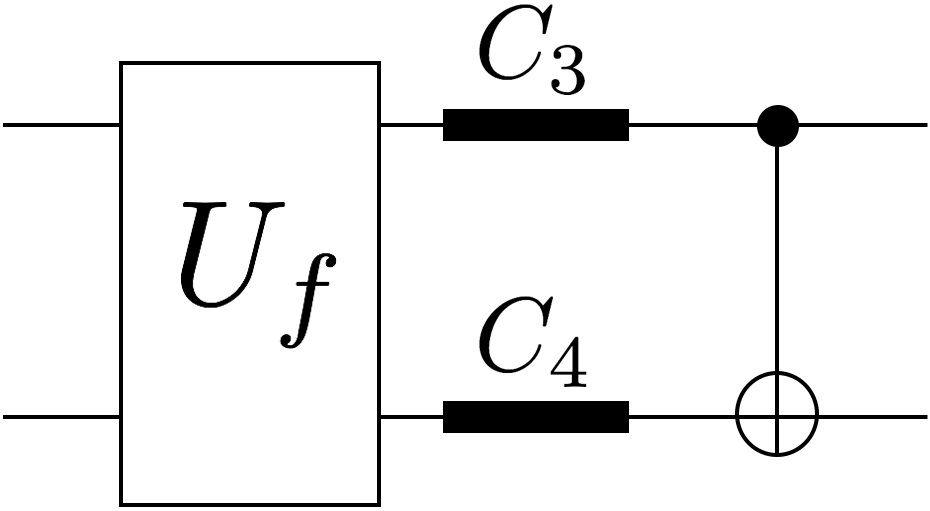}
\caption{}
\label{fig:CNOTtensorProp2}
\end{subfigure}
\caption{(a) Propagating input Clifford gates $C_{1} \otimes C_{2}$ across a CNOT (part of a quantum algorithm) leads to two possible outcomes, one in $C_{good}$ (defined in \cref{eq:cgoodDef}) and the other in $C_{bad}$ (defined in \cref{eq:cbadDef}). (b) Buffers are introduced to learn if the outcome in \cref{fig:CNOTtensorProp1} belongs to $C_{good}$ or $C_{bad}$. Rounds of error correction in the buffers introduce the corrections $C_{3} \otimes C_{4}$. If the outcome in \cref{fig:CNOTtensorProp1} belongs to $C_{bad}$, we apply a CNOT correction following the buffers. The protocol is repeated until the resulting gate belongs to $C_{good}$. }
\label{fig:CNOTCliffordScheme}
\end{figure}

We first address the propagation of logical Clifford corrections (expressed in tensor product form) through CNOT gates. The goal is to prevent a two-qubit correction from spreading to multiple code blocks in order to avoid performing a generic Clifford correction. After the application of a logical CNOT gate (as part of a quantum algorithm), we can place a buffer before the next logical two-qubit gate (or non-Clifford gate) in order to determine what the Clifford frame was right before the application of the logical CNOT. Note that during the application of the buffer, repeated rounds of error correction are performed to prevent the build-up of a large number of errors, producing additional Clifford gates. 

From \cref{eq:PhaseControl,eq:PhaseTarget,eq:HadamardTarget,eq:HadamardControl}, upon propagating the input Clifford gates $C_{1}\otimes C_{2}$ through a logical CNOT gate part of a quantum algorithm using a Clifford frame, two outcomes are possible. In the first case, we can have
\begin{align}
C\mbox{-}X_{12} (C_{1}\otimes C_{2}) = (C'_{1}\otimes C'_{2}) C\mbox{-}X_{12},
\label{eq:InputCNOT1}
\end{align}
for some Clifford gates $C'_{1}$ and $C'_{2}$. In particular, we define
\begin{align}
C_{good} =  (C_{1}\otimes C_{2}) C\mbox{-}X_{12}.
 \label{eq:cgoodDef} 
 \end{align}
If \cref{eq:InputCNOT1} is not satisfied for $C_{1} \otimes C_{2}$, then the output will belong to the set $C_{bad}$ defined to be
\begin{align}
C_{bad} = \mathcal{P}_{2}^{(2)} \setminus C_{good},
\label{eq:cbadDef}
\end{align}
where $ \mathcal{P}_{2}^{(2)}$ is the two-qubit Clifford group. Performing a computer search, we found that out of the $24^{2}=576$ possible input Clifford gates (expressed in tensor product form), $64$ will satisfy \cref{eq:InputCNOT1}.

After the application of the CNOT gate part of the quantum algorithm, the buffers will introduce the Clifford corrections $C_{3} \otimes C_{4}$ arising from the repeated rounds of error correction. Once the Clifford frame prior to applying the CNOT gate is known, we will be able to determine if the output from the propagation of the Clifford frame belongs to $C_{good}$ or $C_{bad}$. If it belongs to $C_{good}$, then no further operations are required. In the case where it belongs to $C_{bad}$, we perform a logical CNOT correction after the buffer. A second set of buffers is introduced to determine if the resulting gate belongs to $C_{good}$ or $C_{bad}$. We can repeat this process recursively until the resulting gate belongs to $C_{good}$.

Assuming the input Clifford gates and buffer Clifford corrections are chosen uniformly at random, we performed a simulation to estimate the transition probability from $C_{bad} \to C_{good}$. Performing $5 \times 10^{5}$ simulations, we found that the $C_{bad}\rightarrow C_{good}$ transition occurs with probability $\frac 1{12}$. Thus, when computing in a random Clifford frame, each logical CNOT requires on average $12+1$ logical CNOTs.

Lastly, we point out that for noise models where the Clifford corrections arising from the buffer are biased towards the Pauli gates, it would be more advantageous to apply $C^{\dagger}_{1} \otimes C^{\dagger}_{2}$ (where $C_{1}$ and $C_{2}$ are the input Clifford gates in \cref{fig:CNOTtensorProp1}) after the following initial CNOT correction. More specifically, if the probability of acquiring a non-trivial Clifford correction in the buffer is $\epsilon$, then the $C_{bad}\rightarrow C_{good}$ transition probability becomes $1-\epsilon\frac{11}{12}$. 

\subsection{Clifford propagation through $T$ gates}
\label{subsubsec:CliffPropT}

We now address the propagation of Clifford corrections through a logical $T$ gate, which is a non-Clifford gate. When applying a logical $T$ gate in a quantum algorithm, we could also place a buffer before the next logical gate part of the quantum algorithm to learn the Clifford frame immediately before applying the $T$ gate. If the output correction is non-Clifford, we can perform appropriate corrections in order to restore the Clifford frame. If successful, this would guarantee that the input correction to the next gate would belong to the Clifford group.

Suppose the input to the logical $T$ gate is a Clifford correction $C_{1}$, so the resulting gate is $TC_1$. On the one hand, if $TC_1 = \tilde C_1T$ for some Clifford $\tilde C_1$---or equivalently  $TCT^\dagger \in \mathcal{P}^{(2)}_{1}$---then no further operations are necessary. Only gates in the set generated by $C_{1} \in \langle S,X \rangle$ satisfy this property. In fact, if $C_{1} \in \langle S,X \rangle$, then $TC_{1}T^{\dagger} \in \langle S,X \rangle$. Thus, we define
\begin{align}
C_{-} = \langle S,X \rangle,\ \ {\rm and}
\ \ C_{+} = \mathcal{P}_{1}^{(2)} \setminus C_{-}.
\label{eq:MinusCliffordSet}
\end{align} 
Note that $|C_{-}| = 8$ while $|C_{+}| = 24-8 = 16$.  Once we learn that the input Clifford correction $C_{1}$ to the logical $T$ gate belongs to $C_{-}$, then no further operations are necessary to restore the Clifford frame. If the buffer Clifford operations are uniformly distributed over the Clifford group, this occurs with probability $\frac 13$.

On the other hand, when we learn that the input Clifford correction $C_{1}$ to the logical $T$ gate belongs to $C_{+}$, we apply another logical $T$ in the hope to restore the Clifford frame. Since an additional Clifford gate $C_2$ was accumulated during the buffer, the resulting gate is $TC_2TC_1$. When $C_2 \in C_-$, then $TC_2TC_1 = \tilde C_2 T^2 C_1 \in \cP_1^{(2)}$, since $T^2 = S$ is a Clifford transformation. At this stage, we have returned to a Clifford frame, but have removed the desired logical $T$ gate: we are thus back at our starting point and can try anew. 

Once again, whenever the Clifford corrections occurring during the buffer are biased to the Pauli group (e.g., when the probability of an error is low), before applying another $T$ gate correction to restore the algorithm $T$ gate, we should apply the Clifford transformation $(\tilde{C}_{2}SC_{1})^{\dagger}$. In this way, we would increase the probability of being in a state of the form $CT$ where $C \in C_{-}$ (thus restoring the Clifford frame). To take this possibility into account, we will henceforth assume that the Clifford corrections arising from the buffer belong to $C_{-}$ with probability $1-p$ and to $C_{+}$  with probability $p$.

Define $\cT^{(0)} = \cP_1^{(2)}$ to be the set of single-qubit Clifford gates, and define 
\begin{equation}
\cT^{(k)} = \prod_{j=1}^k (TC_j)\ \ {\rm where}\ \ C_j\in C_+.
\end{equation}
Every time we learn that the previous buffer Clifford was in $C_+$, we apply a $T$ gate and wait for another buffer. Since this buffer will belong to $C_-$ with probability $1-p$ and to $C_+$ with probability $p$, each step of the above protocol can be seen as a step in a random walk over the sets $\cT^{(k)}$ with transition $\cT^{(k)} \rightarrow \cT^{(k+1)}$ occurring with probability $p$ and transition $\cT^{(k)} \rightarrow \cT^{(k-1)}$ occurring with probability $1-p$. Every time $\cT^{(0)}$ is reached, there is a probability $1-p$ that a logical $T$ gate is successfully realized at the next step. 

To summarize, when attempting to implement a logical $T$ gate starting in a Clifford frame $\cT^{(0)}$, we either succeed with probability $1-p$ or end up applying a $\cT^{(1)}$ gate with probability $p$. In the latter case, we enter a random walk over $k \in  \mathbb{N}$, and our goal it to return to $k=0$. This clearly requires an odd number of steps. The one-step process $1\rightarrow 0$ occurs with probability $1-p$. The three step process $1\rightarrow 2\rightarrow 1 \rightarrow 0$ occurs with probability $p(1-p)^2$. Generalizing to $t=2j+1$ time steps, where $j \in  \mathbb{N}$, the number of paths that lead to a gate in $\mathcal{T}^{(0)}$ is given by the $j$'th Catalan number $K_{j} = \frac{1}{j+1}\binom{2j}{j}$ \cite{KT08}. Therefore, the probability of returning to a gate in $\mathcal{T}^{(0)}$ after $t=2j+1$ time steps is given by
\begin{align}
P_{2j+1} = \frac{1}{j+1} \binom{2j}{j} p^{j} (1-p)^{j+1}.
\label{eq:ProbCatalan}
\end{align}

Using the generating function for the Catalan number $c(x) = \sum_{j \ge 0}K_{j}x^{j} = (1-\sqrt{1-4x})/(2x)$, the total probability that the random walk terminates is given by 
\begin{align}
\sum_{k \ge 0} P_{2k+1} = \min\left\{ \frac{1-p}{p},1\right\}.
\label{eq:ProbHit0}
\end{align}
If $p > 1/2$, then with finite probability the random walk will not terminate whereas if $p \le 1/2$, the random walk is guaranteed terminate. This means that we must choose Clifford gates from $C_{-}$ with probability greater than $1/2$. The latter condition can be satisfied in cases where Clifford corrections arising from the buffer are biased towards gates belonging to $C_{-}$, or in particular if they are biased towards Pauli gates or the identity. When the buffers are chosen uniformly over the Clifford group, then $p=16/24 = 2/3 > 1/2$ so with some finite probability the procedure will not terminate.

\begin{figure}
\centering
\includegraphics[width=0.5\textwidth]{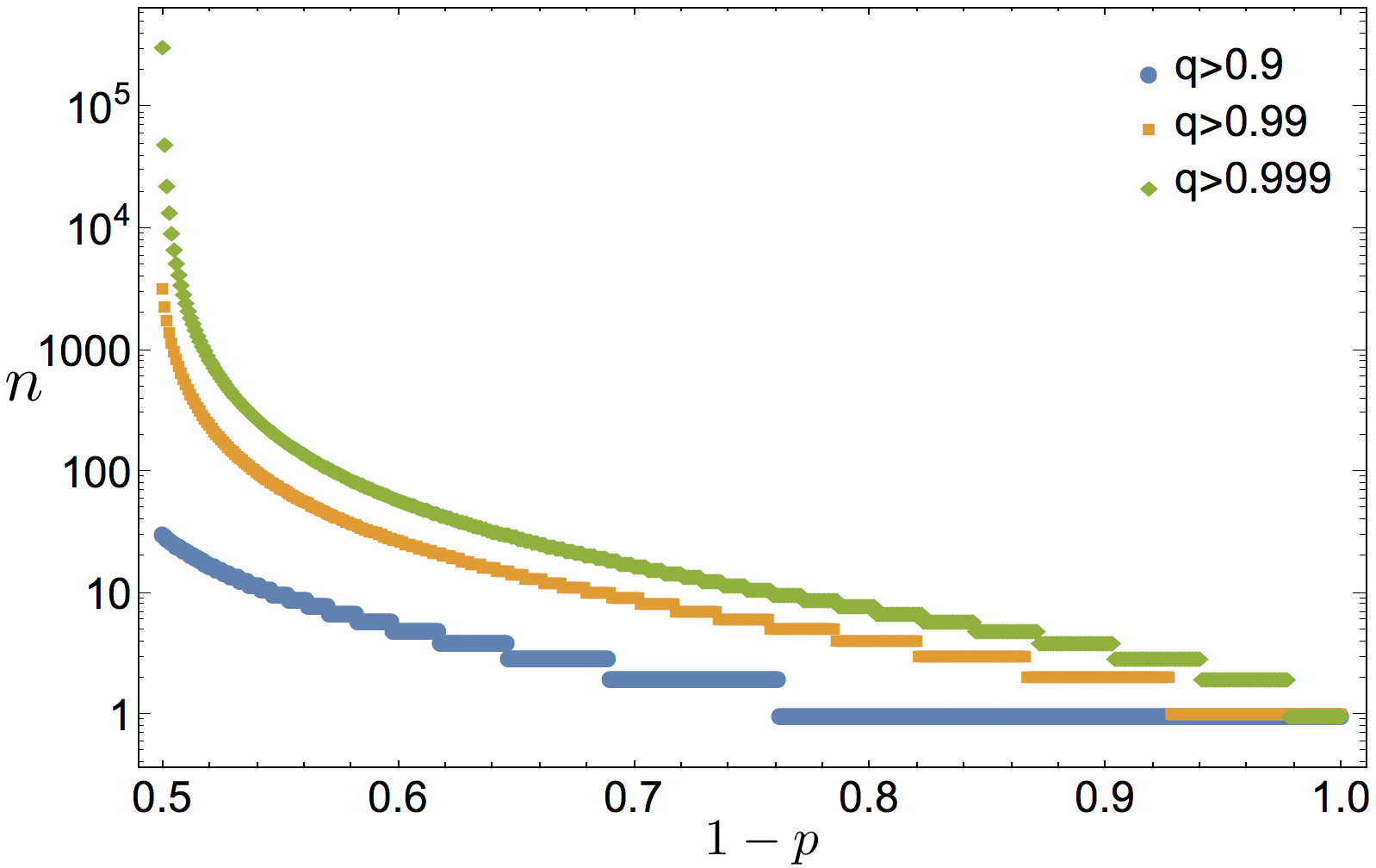}
\caption{For a fixed value of $1-p$, we plot the value of $n$ such that $F(p,n) > q$. We give plots for $q=0.9$, $q=0.99$ and $q=0.999$. Hence, each curve corresponds to the expected number of $T$ gate corrections that are required for obtaining a gate in $\mathcal{T}^{(0)}$ with probability greater than $q$.}
\label{fig:TRemoveScheme}
\end{figure}

We conclude this section by calculating the probability of obtaining a gate in $\mathcal{T}^{(0)}$  within $n$ time steps, which we define as $F(p,n)$. This quantity gives the number of expected $T$ gate corrections that need to be applied in order to restore the Clifford frame after propagation through a logical $T$ gate.  The probability $F(p,n)$ is obtained by summing \cref{eq:ProbCatalan} with a cut-off of $n$ time-steps. The result is given by
\begin{align}
F(p,n) &= \sum_{k = 0}^{n} P_{2k+1} \nonumber \\
&= 1-f(p,n),
\label{eq:CuttOffn}
\end{align}
where
\begin{align}
f(p,n) &= \frac{1-p}{2+n} \binom{2(n+1)}{n+1}(p(1-p))^{n+1} \times \nonumber \\
& \times \ _{2}F_{1}(1,\frac{3}{2}+n;3+n;4p(1-p)),
\label{eq:fDef}
\end{align}
and $_{2}F_{1}(a,b;c;z)$ is the Hypergeometric function defined in \cite{AAR99}.
Plots of \cref{eq:CuttOffn} are given in \cref{fig:TRemoveScheme}.

We now obtain an upper bound on the number of time steps $n$ that are necessary to restore the Clifford frame with probability $q = 1 - \epsilon$. In other words, we would like to obtain an upper bound on $n$ such that $F(p,n) > 1 - \epsilon$. We first obtain a lower bound for the function $f(p,n)$ by using the following inequalities
\begin{align}
\binom{2(n+1)}{n+1} \ge \Bigg (\frac{2(n+1)}{(n+1)} \Bigg )^{n+1} = 2^{n+1},
\label{eq:BinomialInequality}
\end{align}
\begin{align}
_{2}F_{1}(1,\frac{3}{2}+n;3+n;4p(1-p)) \ge 2p+1.
\label{eq:HypInequality}
\end{align}
Inserting \cref{eq:BinomialInequality,eq:HypInequality} into \cref{eq:CuttOffn}, we obtain
\begin{align}
F(p,n) \le 1 - \frac{(1-p)(2p+1)}{n+2} [2p(1-p)]^{n+1}.
\label{eq:UpperBoundFpn}
\end{align}
Setting $F(p,n) > 1 - \epsilon$, we obtain 
\begin{align}
\frac{[2p(1-p)]^{n+1}}{n+2} \le \frac{\epsilon}{(1-p)(2p+1)}
\label{eq:IntermediateStep}
\end{align}
For $n \ge 1$, we have $1/(n+2) > (\frac 12)^{n+1}$, so \cref{eq:IntermediateStep} becomes
\begin{align}
[p(1-p)]^{n+1} \le \frac{\epsilon}{(1-p)(2p+1)}
\label{eq:IntermediateStep2}
\end{align}
which shows that $n \sim |\log{\epsilon}|$---the $T$ gate overhead for restoring the Clifford frame scales logarithmically in $\epsilon$.

We now consider regimes where the noise is below the fault-tolerance threshold of a code (say $p \lesssim 10^{-2}$ as is required for the surface code \cite{FMMC12}). In such regimes, corrections based on syndrome measurement outcomes will be significantly biased towards the identity, or more generally towards Pauli operators. More specifically, suppose that for a given noise model and code, a Pauli correction leads to a logical error rate $\delta_P$.  Applying Clifford corrections can only improve this logical error rate to $\delta_C \leq \delta_P$ since they include Pauli corrections. But in a Clifford correction scheme, non-Pauli corrections are only used when Pauli corrections are not optimal, which occurs at a rate $\delta_P$, which is small below threshold.  Consequently, in the case where a correction from $C_{bad}$ was applied, the expected number of $T$ gate corrections to return to the Clifford frame (as shown in \cref{fig:TRemoveScheme}) would be very close to one since $p\lesssim \delta_P$. Thus, we conclude that the Clifford gates can be used with the scheme proposed here at a negligible cost.

We conclude this section by mentioning that Clifford frames are also useful in randomized benchmarking schemes where random Clifford gates are applied to transform a given noise channel into an effective depolarizing channel \cite{MGE11PRL,MGE12PRA,WE16}. In these schemes, once the random Clifford gates have been applied, they must be propagated in classical software through a sequence of logical gates that are part of the quantum circuit of interest (which typically includes Clifford and $T$ gates) and conjugate Clifford gates are subsequently applied after the Clifford frame has been restored. The techniques presented in this section can be used to restore the Clifford frame after propagation through the logical gates and could thus be an attractive approach for performing randomized benchmarking using Clifford gates. Note that if only random Pauli operators were used, the noise would be transformed to a Pauli channel (but not in general a depolarizing channel).

\section{Conclusion}

In this paper, we considered performing fault-tolerant quantum computation when measurement times and/or decoding times are much slower than gate times. 

This was realized by providing a new scheme for performing error correction using the Pauli frame by placing buffers after the application of a non-Clifford gate. We showed that the Pauli frame can always be restored by applying logical Clifford corrections prior to the application of the next two-qubit or non-Clifford gate part of a quantum algorithm.

Given that logical Clifford corrections can significantly increase a code's threshold value \cite{CWBL16}, in the remainder of the manuscript, we showed how to perform fault-tolerant error correction in the Clifford frame. We performed an in-depth study of the propagation of logical Clifford corrections through logical CNOT and $T$ gates, and the same idea can be generalized to other universal gate sets. 

For the propagation through CNOT gates, we placed buffers at strategic locations to ensure that the output Clifford corrections could be expressed in tensor product form. To achieve this, we found that on average 12 logical CNOT corrections would be required when Clifford corrections arising from buffers are chosen uniformly at random. This ensures that two-qubit Clifford corrections would not propagate through the remainder of the circuit yielding a generic Clifford correction.

We also used buffers to keep track of the Clifford corrections propagating through $T$ gates.  We showed that in certain conditions, by applying enough $T$ gate corrections, the Clifford frame could be restored with probability arbitrarily close to 1. Furthermore, to restore the Clifford frame with probability at least $1- \epsilon$, we showed that the number of $T$ gate corrections scales as $\log{(1/\epsilon})$.

While \cite{CWBL16} has shown that Clifford corrections can produce a higher error threshold, the impact and applicability of Clifford corrections remains largely unexplored. Our original motivation for the current work was to determine if one of the key tricks used in FT schemes --- Pauli frames --- could be generalized to Clifford corrections. Having found that it can indeed be generalized at a negligible cost below threshold, we pave the way to future investigations of Clifford corrections. 

\section{Acknowledgements}
We thank Tomas Jochym-O'Connor and Daniel Gottesman for useful discussions. C. C. would like to acknowledge the support of QEII-GSST and to thank the Institut Quantique at the University of Sherbrooke for their hospitality where most of this work was completed. This work was supported by the Army Research Office contract number W911NF-14-C-0048.

\bibliographystyle{unsrtnat} 
\bibliography{bibtex_chamberland}

\clearpage
\appendix

\section{Error correction for input Clifford errors}
\label{app:CliffordAndPureError}
In this appendix we will explain why a generic $n-$qubit Clifford error on an encoded state transforms into a logical Clifford operation and a physical Pauli operation, upon doing stabilizer measurements.

Before proceeding, we provide a few definitions. Let $\bar{\rho}$ be an encoded state of a stabilizer code with stabilizer $\cS$, that undergoes an error described by some CPTP map $\cE$, i.e, $\bar{\rho} \mapsto \cE(\bar{\rho})$. Let $\Pi_{0}$ denote the projector onto the code space and consequently $\Pi_{s}$ denote the projector onto the syndrome space of $s$. Let $T_{s}$ be a Pauli operation that takes a state from the syndrome $s$ subspace to the code space, i.e, $\Pi_{s} = T_{s}\cdot \Pi_{0} \cdot T_{s}$. Upon measuring the stabilizer generators to obtain a syndrome $s$ and applying the corresponding $T_{s}$ operation, the noisy state can be projected back to the code space,
\begin{gather}
\cE(\bar{\rho}) \mapsto T_{s}\Pi_{s} \thickspace \cE(\bar{\rho}) \thickspace \Pi_{s} T_{s} \thickspace . \label{noiseProjection}
\end{gather}
Since the resulting state is in the codespace, it can be decoded back to a single qubit state $\phi$, given by
\begin{gather}
\phi = \sum_{a} \trace{T_{s}\Pi_{s} \thickspace \cE(\bar{\rho}) \thickspace \Pi_{s} T_{s} \cdot \bar{P}_{a}} P_{a} \thickspace \label{decodedState}
\end{gather}
Let us denote the combined effect of the encoder, noise model, the projection to code space and the decoder by a quantum channel called the \emph{effective projection}. The effective projection can be seen as acting directly on $\rho$ to result in $\phi$ \cite{RDM02}.

However, in order to extract the (single qubit or logical) CPTP map defining the effective projection channel, we must make use of another tool, an isomorphism between channels and states, called the Choi-Jami\l{}owski isomorphism \cite{AJ72,WCJ15}. Under this isomorphism, a single qubit CPTP channel $\cE$ is mapped to a unique bipartite quantum state $\cJ(\cE)$ called the \emph{Choi matrix} corresponding to $\cE$, in the following manner.
\begin{gather}
\cJ(\cE) = \dfrac{1}{4}\sum_{i,j}\cE(P_{i}) \otimes P^{T}_{j} \label{choiChannel}
\end{gather}
Furthermore, when the quantum channel is expressed as a process matrix $\Lambda$, where $\Lambda_{i,j} = \trace{\cE(P_{i}) \cdot P_{j}}$, where $P_{i}, P_{j} \in \{I, X, Y, Z\}$ are Pauli matrices, it follows that
\begin{gather}
\Lambda_{ij} = \trace{\cJ(\cE) \cdot (P_{j}\otimes P^{T}_{i})} \label{stateToChannel} \thickspace .
\end{gather}

Lastly, note that when $\cE(\rho) = C\cdot \rho\cdot C^{\dagger}$ where $C$ is a Clifford operation, the corresponding process matrix $\Lambda(\cE)$ is given by
\begin{flalign}
\Lambda(\cE)_{ij} &= \trace{C\cdot P_{i}\cdot C^{\dagger} \cdot P_{j}} \nonumber \\
&= \trace{P_{\sigma(i)} \cdot P_{j}} = \delta_{\sigma(i), j} \label{processCliffordChannel}
\end{flalign}
where $\sigma$ is a permutation of $\{1, 2, 3, 4\}$ that depends on the Clifford operation. Hence the process matrix of a Clifford channel is a permutation matrix.

We now have all the necessary ingredients to prove our claim. Note that if $\rho$ in Eq. \ref{noiseProjection} is the Bell state and $\cE$ acts on one half of the encoded Bell state, then using Eq. \ref{choiChannel}, we know that $\phi$ in Eq. \ref{decodedState} is simply the Choi matrix corresponding to the effective projection channel. From Eq. \ref{stateToChannel}, it follows that the process matrix for the effective projection channel, denoted by $\Lambda^{(1)}$, is given by
\begin{flalign}
\Lambda^{(1)}_{ij} &= \trace{T_{s}\Pi_{s} \thickspace \cE(\Pi_{0} \cdot \bar{P}_{i}) \thickspace \Pi_{s} T_{s} \Pi_{0} \bar{P}_{j}} \nonumber \\
&= \trace{\cE(\Pi_{0}\cdot \bar{P}_{i}) \cdot \Pi_{s} \cdot \bar{P}_{j}} \nonumber \\
&= \sum_{\substack{l \\ P_{l} \in \bar{P}_{i}\cdot \cS}}\sum_{\substack{k \\ P_{k} \in \bar{P}_{j}\cdot \cS}} c_{k}\Lambda_{lk} \nonumber \\
&= \sum_{\substack{l \\ P_{l} \in \bar{P}_{i}\cdot \cS}}\sum_{\substack{k \\ P_{k} \in \bar{P}_{j}\cdot \cS}} c_{k}\delta_{l, \sigma(k)} \label{cliffProjectionChannel}
\end{flalign}
where $c_{k} \in \{+1, -1\}$ and in the last step, we have used the fact that $\Lambda$ is the process matrix of a Clifford operation, Eq. \ref{processCliffordChannel}. That $\Lambda^{(1)}$ is also a permutation matrix follows from two properties -- (i) Every row of $\Lambda^{(1)}$ has a unique non-zero column and (ii) every column of $\Lambda^{(1)}$ has a unique non-zero row. We will show (i) explicitly in what follows, the proof for (ii) is almost identical. Suppose that there are two columns $j^{\prime}$ and $j^{\prime\prime}$ such that $\Lambda_{i j^{\prime}} \neq 0$ and $\Lambda_{i j^{\prime\prime}} \neq 0$. Along with Eq. \ref{cliffProjectionChannel}, it implies that there exists $P_{k^{\prime}} \in \bar{P}_{j^{\prime}}\cdot \mathcal{S}$ and $P_{k^{\prime\prime}} \in \bar{P}_{j^{\prime\prime}}\cdot \mathcal{S}$ such that $\delta_{l, \sigma(k^{\prime})} = \delta_{l, \sigma(k^{\prime\prime})}$. Hence, it must be that $\sigma(k^{\prime}) = \sigma(k^{\prime\prime})$, in other words, $j^{\prime} = j^{\prime\prime}$.

Hence the effective projection channel is indeed a Clifford operation, in other words, any $n-$qubit Clifford operation on the physical qubits can be promoted to a logical Clifford operation and a physical Pauli error ($T_{s}$ in Eq. \ref{noiseProjection}), by a syndrome measurement.

\end{document}